\documentclass[showpacs,preprintnumbers,amsmath,amssymb,twocolumn]{revtex4}%

\usepackage{graphicx}% Include figure files
\usepackage{dcolumn}% Align table columns on decimal point
\usepackage{bm}% bold math
\usepackage{color}

\begin{document}

\title{Particle motion and collisions around rotating regular black hole}

\author{Bobir Toshmatov$^{1}$}
\email{b.a.toshmatov@gmail.com}

\author{Ahmadjon Abdujabbarov$^{1,2,3}$}
\email{ahmadjon@astrin.uz}

\author{Bobomurat Ahmedov$^{1,2}$}
\email{ahmedov@astrin.uz}

\author{Zden\v{e}k Stuchl\'{i}k$^{3}$}
\email{zdenek.stuchlik@fpf.slu.cz}

\affiliation{%
$^{1}$ Institute of Nuclear Physics, Ulughbek, Tashkent 100214, Uzbekistan\\
$^{2}$ Ulugh Beg Astronomical Institute, Astronomicheskaya 33, Tashkent 100052, Uzbekistan\\
$^{3}$ Faculty of Philosophy \& Science, Institute of Physics, Silesian University in Opava,
Bezru\v{c}ovo n\'{a}m\v{e}st\'{i} 13, CZ-74601 Opava, Czech Republic}

\begin{abstract}
The neutral particle motion around rotating regular
%TAAS (Toshmatov-Ahmedov-Abdujabbarov-Stuchl\'{i}k)
black hole that was derived from the Ay\'{o}n-Beato-Garc\'{i}a
black hole solution by the Newman-Janis algorithm in the preceding
paper [Phys. Rev. D 89, 104017, (2014)] has been studied. The
dependencies of the ISCO (innermost stable circular orbits along
geodesics) and unstable orbits on the value of the electric charge
of the rotating regular black hole have been shown. Energy
extraction from the rotating regular black hole through various
processes has been examined. We have found expression of the
center of mass energy for the colliding neutral particles coming
from infinity, based on the BSW (Ba\v{n}ados-Silk-West) mechanism.
{The electric charge $Q$ of rotating regular black hole decreases
the potential of the gravitational field and the particle needs
less bound energy at the circular geodesics. This causes the
increase of efficiency of the energy extraction  through BSW
process in the presence of the electric charge $Q$ from rotating
regular black hole.} It has been shown that the efficiency of the
energy extraction from the rotating regular black hole via the
Penrose process decreases with the increase of the electric charge
$Q$ and is smaller with compare to 20.7~\% which is the value for
the extreme Kerr black hole with the specific angular momentum
$a=1$.

\end{abstract}

\maketitle

\section{Introduction}

Recently, Ba\v{n}ados, Silk and West (BSW)~\cite{Banados}  have
shown that free particles falling from rest at infinity outside a
Kerr black holes may collide with arbitrarily high center-of-mass
(CM) energy and hence the maximally rotating black hole with the
specific angular momentum $a=1$ might be regarded as a
Planck-energy-scale collider. They have proposed that this might
lead to observable signals from the ultra high energy collisions
of the particles around the black holes.

One peculiar feature of string theory, which may play a role of
one of the candidates for a theory of quantum gravity, is the
presence of extra dimensions of space. The existence of the extra
dimensions amplifies the gravity of the central object
significantly. For example, in the preceding paper~\cite{Arman} we
have studied particle acceleration around black strings in
$S^2\times R^1$ topology which are produced based on the string
theory by adding extra dimension to the Schwarzschild and Kerr
black hole spacetimes. In the papers~\cite{Mandar,ss10} Kerr naked
singularities and superspinars have been studied as  particle
accelerators and shown that the center-of-mass energy of the
collision between two particles is arbitrarily large in the near
extremal Kerr naked singularity and superspinar. The evolution of Kerr superspinars due to accretion counterrotating thin discs have been studied in~\cite{Stuchlik11}. Ultra-high-energy collisions of particles in the field of near-extreme Kehagias-Sfetsos naked singularities have been considered in~\cite{Stuchlik14}

Near to extreme rotation is essential to achieve the unlimited
energy for the center of the mass through the particle
acceleration around rotating black holes. However, in the
paper~\cite{Frolov} it has been shown that static black hole can
be also particle accelerator when the black hole is immersed in
the external magnetic field.

It is well known that there are so-called regular black holes
which do not have  curvature singularity, see
e.g.~\cite{Bobir,Eloy1} and~\cite{Eloy}. After derivation of the
Kerr solution from the Schwarzschild one by applying the
Newman-Janis algorithm (NJA)~\cite{Drake}, this algorithm has been
widely used to get rotational solutions of black holes, see
e.g.~\cite{Bambi}. In our preceding research~\cite{Bobir} the
rotational solution
%(abbreviated as TAAS from the names of authors
%Toshmatov-Ahmedov-Abdujabbarov-Stuchl\'{i}k)
of the
Ay\'{o}n-Beato-Garc\'{i}a static regular black hole~\cite{Garcia} one has been
found by using the Newman-Janis algorithm. In this paper we study
the particle motion around the newly derived rotating regular
black hole as well as the efficiency of the energy extraction by
the Penrose and BSW processes.

The paper is organized in the following way. In the
Sec.~\ref{spacetime} we study the effective potential and types of
particle orbits around rotating regular black hole. The innermost
stable circular geodesics around rotating regular black hole are
discussed in the Sec.~\ref{isco}. The next two Secs.~\ref{bsw}
and~\ref{penrose} are devoted to the energy extraction from the
rotating regular black hole through BSW mechanism and Penrose
process, respectively. In the Sec.~\ref{conclusion} we summarize
our main results.

In the paper, we use a spacetime signature as $(-,+,+,+)$ and a
system of geometric units in which $G = 1 = c$. Greek indices are
taken to run from 0 to 3.

\section{The Particle Orbits around Rotating Regular Black Hole}
\label{spacetime}

\begin{figure}[h]
\centering
\includegraphics[width=8cm]{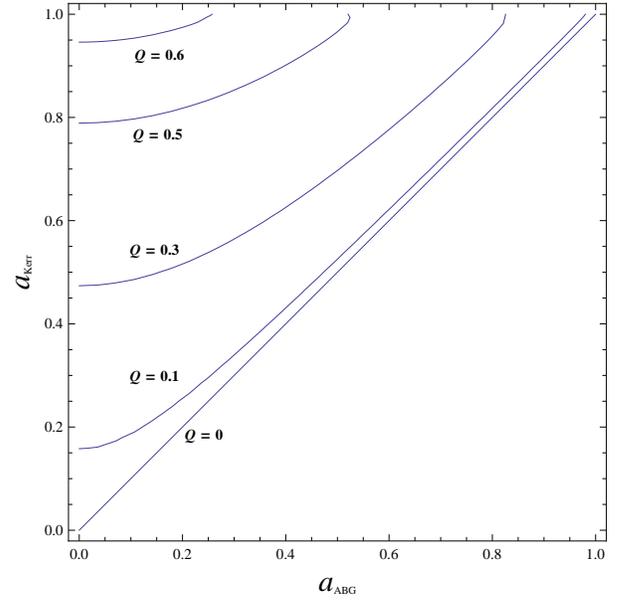}
\caption{\label{rotate} Relationship between the rotation
parameters of the Kerr $a_{Kerr}$ and regular $a_{ABG}$ black
holes for the several values of the electric charge $Q$ when the
radii of their event horizons coincide
($r_{+}^{Kerr}=r_{+}^{ABG}$).}
\end{figure}

The line element in the space-time of the rotating regular black
hole in the Boyer-Lindquist coordinates  is given as \cite{Bobir}
\begin{eqnarray}\label{00}
d s^2&=&g_{tt}dt^2+g_{rr}dr^2+2 g_{t\phi}d\phi dt+g_{\theta\theta}d\theta^2+g_{\phi\phi}d\phi^2,\nonumber\\
\end{eqnarray}
with
\begin{eqnarray}\label{01}
&&g_{tt}=-f(r,{\theta}),\nonumber\\
&&g_{rr}=\frac{\Sigma}{\Sigma f(r,{\theta})+a^2\sin^2\theta},\nonumber\\
&&g_{t\phi}=-a\sin^2\theta(1-f(r,{\theta})),\nonumber\\
&&g_{\theta\theta}=\Sigma,\nonumber\\
&&g_{\phi\phi}=[\Sigma-a^2(f(r,{\theta})-2)\sin^2\theta]\sin^2\theta\
,
\end{eqnarray}
where
\begin{eqnarray}\label{02}
&&f(r,{\theta})=1-\frac{2M r \sqrt{\Sigma}}{(\Sigma+Q^2)^{3/2}}+\frac{Q^2\Sigma}{(\Sigma+Q^2)^2},\\
&&\Sigma=r^2+a^2 \cos^2\theta \ .
\end{eqnarray}
{where $M$, $a$ and $Q$ are the total mass, the specific angular
momentum and the electric charge of the black hole, respectively.}
The space-time metric (\ref{00}) is identical to {the
Ay\'{o}n-Beato-Garc\'{i}a one when the the specific angular
momentum $a=0$~\cite{Garcia},} to the Schwarzschild one when the
specific angular momentum $a=0$, the electric charge $Q=0$ and to
the Kerr black hole when $Q=0$ (see Fig.~\ref{rotate}).

In order to compute the trajectories of the geodesic motion of a
test particle in the equatorial plane ($\theta=\pi/2$) one needs
the Lagrangian for this motion:
\begin{eqnarray}\label{03}
L=\frac{1}{2}[g_{tt}\dot{t}^2+2g_{t\phi}\dot{t}\dot{\phi}+g_{rr}\dot{r}^2+g_{\phi\phi}\dot{\phi}^2]\
,
\end{eqnarray}
where an overdot denotes the derivative with respect to proper
time $\tau$. Since the Lagrangian (\ref{03}) does not depend on
the coordinates $t$ and $\phi$, associated momenta $p_{t}$ and
$p_{\phi}$ are conserved and they are called energy $E$ and
angular momentum $L$ of a test particle, respectively:
\begin{eqnarray}\label{04}
&&p_t=\xi_{(t)}^\mu
u_{\mu}=g_{tt}\dot{t}+g_{t\phi}\dot{\phi}=-\frac{E}{m}=-{\cal E}=
\textrm{const}\ ,
\end{eqnarray}
\begin{eqnarray}\label{05}
&&p_\phi=\xi_{(\phi)}^\mu
u_{\mu}=g_{t\phi}\dot{t}+g_{\phi\phi}\dot{\phi}=\frac{L}{m}={\cal
L}=\textrm{const}\ ,
\end{eqnarray}
\begin{eqnarray}\label{06}
&&p_r=g_{rr}\dot{r}\ ,
\end{eqnarray}
\begin{eqnarray}\label{060}
&&p_\theta=0\ ,
\end{eqnarray}
where $m$ is mass of the test particle.  ${\cal E}$ and ${\cal L}$
are the specific energy and angular momentum of the test particle
per unit mass $m$.

The Hamiltonian is given by ${\cal
H}=p_t\dot{t}+p_r\dot{r}+p_\phi\dot{\phi}-L$. In terms of the
components of the metric tensor the Hamiltonian is
\begin{eqnarray}\label{07}
&&2{\cal H}=(g_{tt}\dot{t}+g_{t\phi}\dot{\phi})\dot{t}+g_{rr}\dot{r}^2+
(g_{t\phi}\dot{t}+g_{\phi\phi}\dot{\phi})\dot{\phi}\nonumber\\
&&=-{\cal E}\dot{t}+{\cal
L}\dot{\phi}+g_{rr}\dot{r}^2=\epsilon=\textrm{const}\ .
\end{eqnarray}
Here parameter $\epsilon$ is equal to either -1, 0 or +1 for
time-like, light-like (null) and spacelike geodesics,
respectively. By solving equations (\ref{04}) and (\ref{05})
simultaneously, one can find
\begin{eqnarray}\label{08}
\dot{t}=\frac{g_{\phi\phi}{\cal E}+g_{t\phi}{\cal
L}}{g_{t\phi}^2-g_{tt}g_{\phi\phi}},\
\dot{\phi}=-\frac{g_{t\phi}{\cal E}+g_{tt}{\cal
L}}{g_{t\phi}^2-g_{tt}g_{\phi\phi}}\ ,
\end{eqnarray}
where, the event horizon $r_{+}$ is given by the  larger root of
denominator of the expressions (\ref{08})
$g_{t\phi}^2-g_{tt}g_{\phi\phi}=0$. It can be easily verified that
$g_{t\phi}^2-g_{tt}g_{\phi\phi}=0$ $\Leftrightarrow$ $\Sigma
f(r)+a^2\sin^2\theta=0$.

By inserting (\ref{08}) into (\ref{07}) and considering particle
as moving along the time-like geodesics ($\epsilon=-1$) one can
obtain the expression for the radial velocity of the particle
around rotating regular black hole as
\begin{eqnarray}\label{005}
&&\dot{r}^2={\cal E}^2-\frac{(Q^2-2\sqrt{r^2+Q^2})(a{\cal E}-{\cal L})^2}{(r^2+Q^2)^2}\nonumber\\
&&+\frac{a^2{\cal E}^2-{\cal
L}^2}{r^2}-\frac{r^2(Q^2-2\sqrt{r^2+Q^2})}{(r^2+Q^2)^2}-\frac{a^2}{r^2}-1\
,
\end{eqnarray}

By introducing the notion of effective potential
\begin{eqnarray}\label{veff}
&&V_{eff}=\frac{(Q^2-2\sqrt{r^2+Q^2})(a{\cal E}-{\cal L})^2}{(r^2+Q^2)^2}-\frac{a^2{\cal E}^2-{\cal L}^2}{r^2}\nonumber\\
&&+\frac{r^2(Q^2-2\sqrt{r^2+Q^2})}{(r^2+Q^2)^2}+\frac{a^2}{r^2}+1\
,
\end{eqnarray}
one can write~(\ref{005}) as
\begin{eqnarray}\label{006}
\dot{r}^2={\cal E}^2-V_{eff}\ .
\end{eqnarray}

\begin{figure*}[t!.]
\begin{center}
\includegraphics[width=0.45\linewidth]{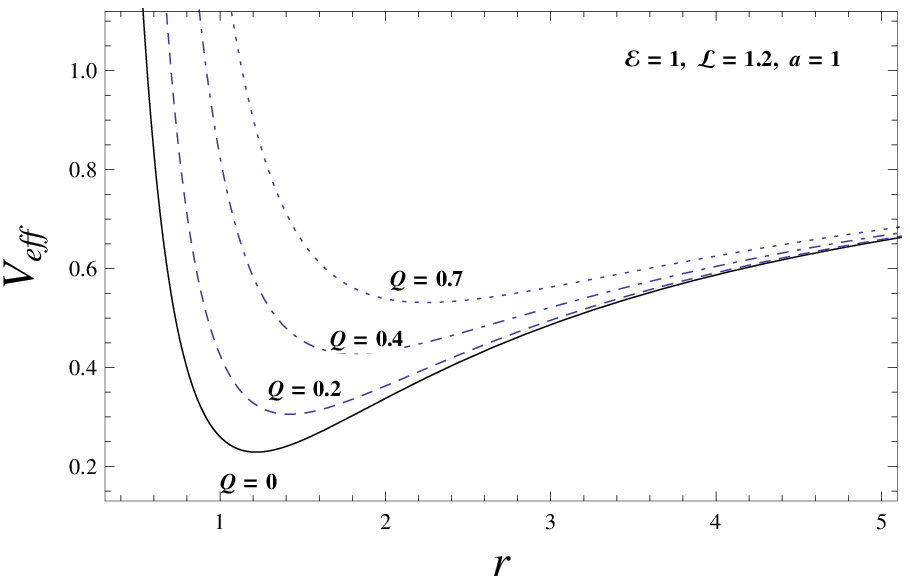}
\includegraphics[width=0.45\linewidth]{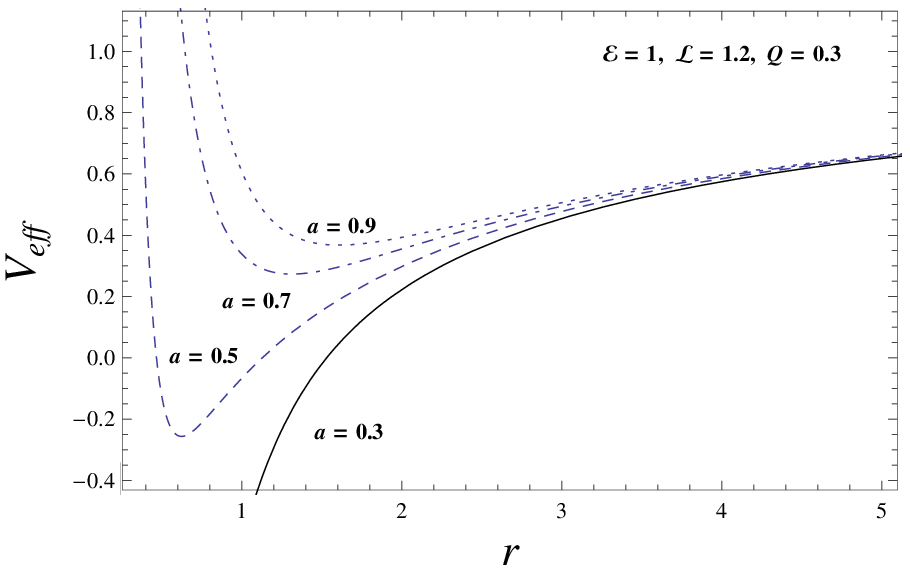}
\end{center}
\caption{\label{fig_bobir} The radial  dependence of the effective
potential of the particle moving around rotating regular black
hole for the different typical values of the electric charge $Q$
(left panel) and the rotation parameter $a$ (right panel). $Q=0$
(solid line in the left panel) corresponds to the effective
potential of the Kerr black hole.}
\end{figure*}

One can see from the expression (\ref{veff}) for the effective
potential that the motion of particle around rotating regular
black hole is invariant under $r\longleftrightarrow-r$ and
$Q\longleftrightarrow-Q$ transformations. In the flat spacetime
limit ($r\rightarrow\infty$) effective potential of the motion of
the particle $V_{eff}$ tends to $1$ ($V_{eff}\rightarrow1$). In
the opposite limiting case $r\rightarrow0$, effective potential of
the motion of the particle $V_{eff}$ tends to infinity
($V_{eff}\rightarrow\infty$).

As it has been shown in our preceding research~\cite{Bobir} with
the increase of the value of the electric charge $Q$ the horizon
of the   black hole decreases and eventually, in the value of the
charge $Q>0.633$ event horizon vanishes.

\begin{figure*}[t!.]
\begin{center}
\includegraphics[width=0.32\linewidth]{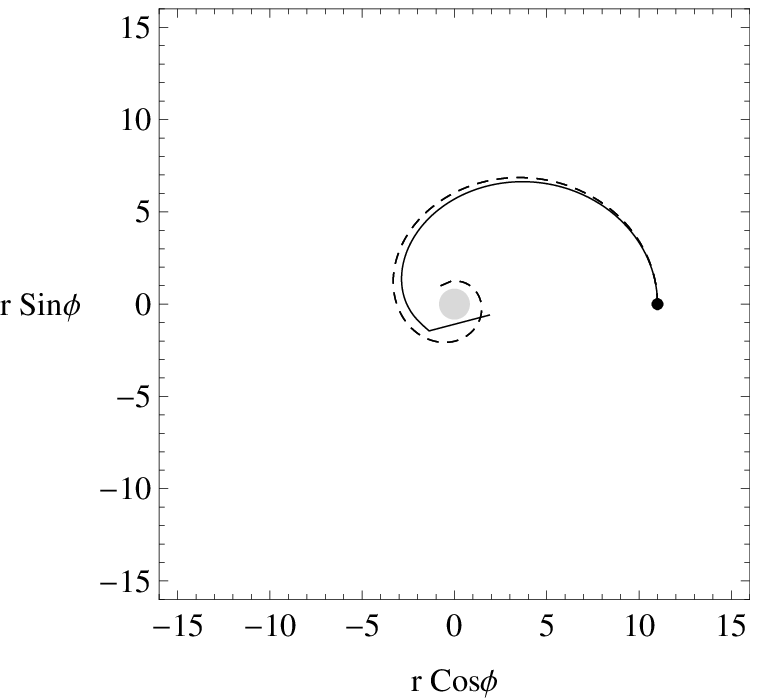}
\includegraphics[width=0.32\linewidth]{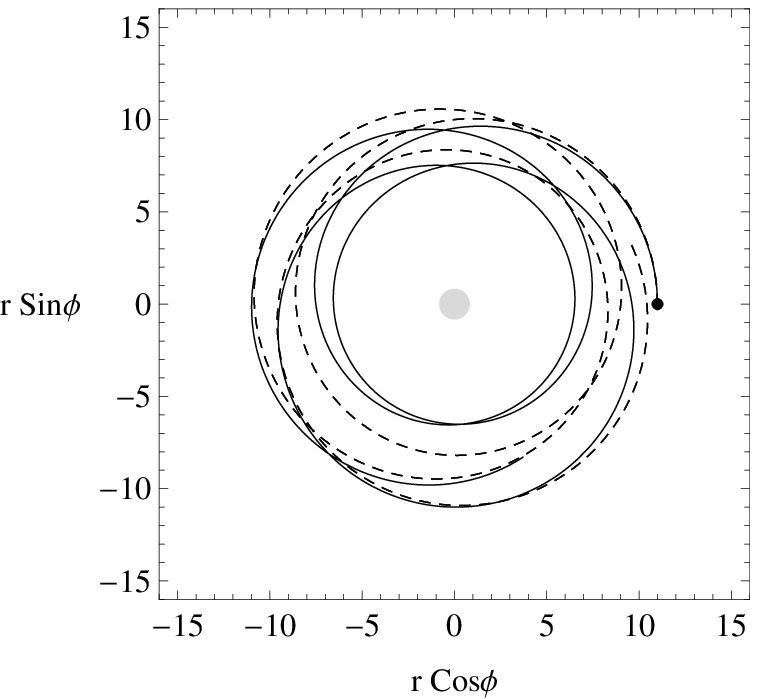}
\includegraphics[width=0.32\linewidth]{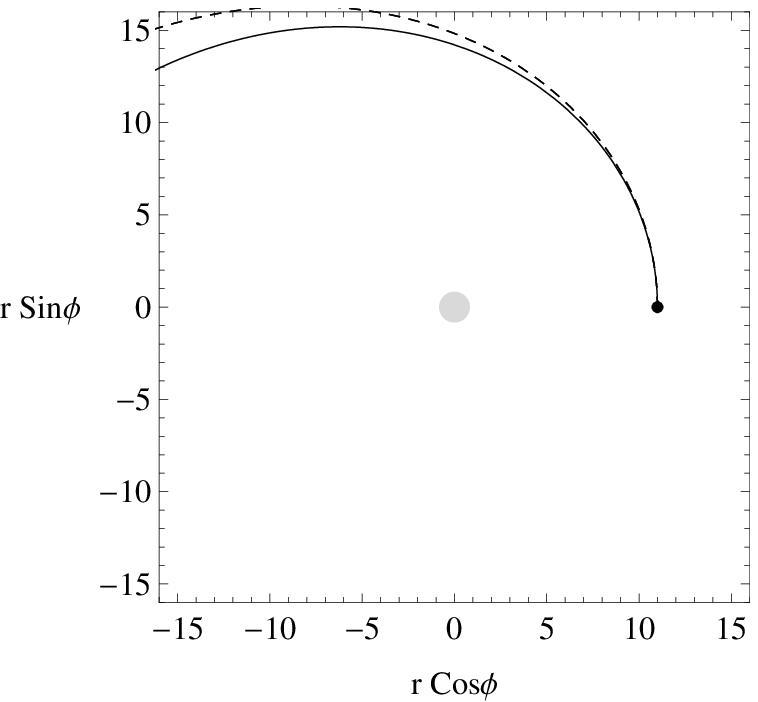}
\end{center}
\caption{\label{fig_bobir1} Examples of particle trajectories
moving in the equatorial plane ($\theta=\pi/2$) around Kerr
(solid, $Q=0$) and rotating regular   (dashed, $Q=0.6$) BHs in the
presence of the event horizon of   BH when the rotation parameter
$a=0.2$: terminating orbit, bound orbit and escape orbit (from
left to right). Particle starts motion from the initial position
$r_{0}=11$ with the different values of the specific angular
momentum ${\cal L}=2.6$, ${\cal L}=3.5$ and ${\cal L}=4.5$.}
\end{figure*}

\begin{figure*}[t!.]
\begin{center}
\includegraphics[width=0.32\linewidth]{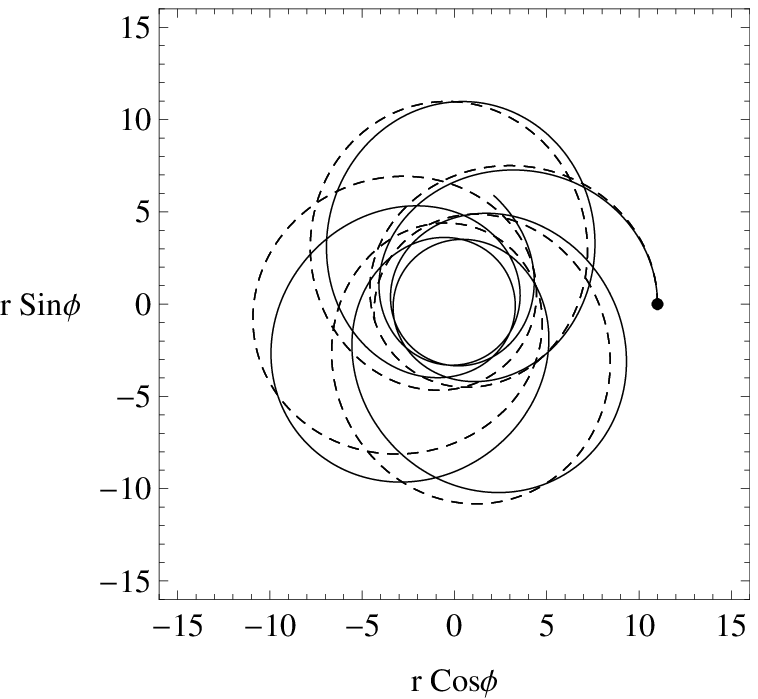}
\includegraphics[width=0.32\linewidth]{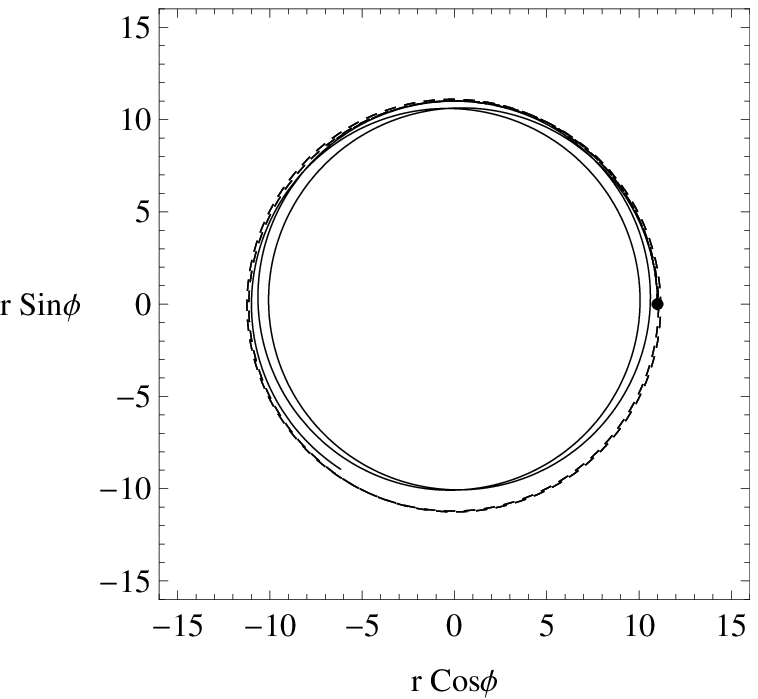}
\includegraphics[width=0.32\linewidth]{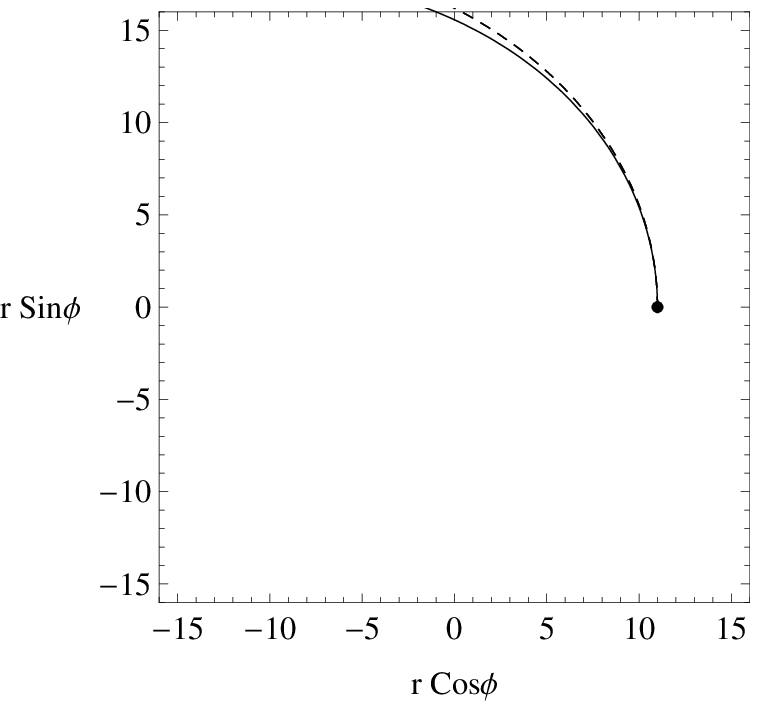}
\end{center}
\caption{\label{fig_bobir2} Examples of particle trajectories
moving in the equatorial plane ($\theta=\pi/2$) around Kerr
(solid, $Q=0$) and rotating regular   (dashed, $Q=0.6$) BHs in the
absence of the event horizon of   BH when the rotation parameter
$a=0.99$. Particle starts motion from the initial position
$r_{0}=11$ with the different values of the specific angular
momentum ${\cal L}=2.6$, ${\cal L}=3.5$ and ${\cal L}=4.5$. }
\end{figure*}

It is known that there are  three types of particle orbits around
central compact gravitating object: \textit{terminating orbit},
\textit{bound orbit} and \textit{escape orbit}. These orbits are
characterized by the angular momentum ${\cal L}$ of the particle.
In Figs.~\ref{fig_bobir1} and~\ref{fig_bobir2} examples of the
particle orbits around   black hole are given in the cases of
presence and absence of the horizon of the black hole,
respectively.

\section{Innermost Stable Circular Orbits}
\label{isco}

From the astrophysical point of view one of the most  momentous
type of orbits of the particle is innermost stable circular orbit
(ISCO). The ISCO can be found by solving the second derivative of
the effective potential $V_{eff}$ with respect to the radial
coordinate $r$, i.e.
\begin{eqnarray}\label{09}
\frac{d^2V_{eff}}{dr^2}=0\ .
\end{eqnarray}
It is known that the effective potential of the motion
(\ref{veff})  is the function of the specific energy ${\cal E}$
and angular momentum ${\cal L}$ as well. One can find the energy
${\cal E}$ and angular momentum ${\cal L}$ from the following
equations:
\begin{eqnarray}\label{010}
\dot{r}=0, && \frac{dV_{eff}}{dr}=0\ .
\end{eqnarray}
The first equation of (\ref{010}) is responsible for orbits  being
circular, namely it indicates the turning point of the path of the
motion. If particle moves along the circular geodesics the motion
of the particle will be finite. The range of radius $r$ of the
motion is found from the second equation of (\ref{010}). Solving
the equations (\ref{010}) with respect to ${\cal E}$ and ${\cal
L}$, simultaneously, and inserting produced expressions into
(\ref{09}) one can find the region of the ISCO.
\begin{figure*}[t!.]
\begin{center}
\includegraphics[width=0.45\linewidth]{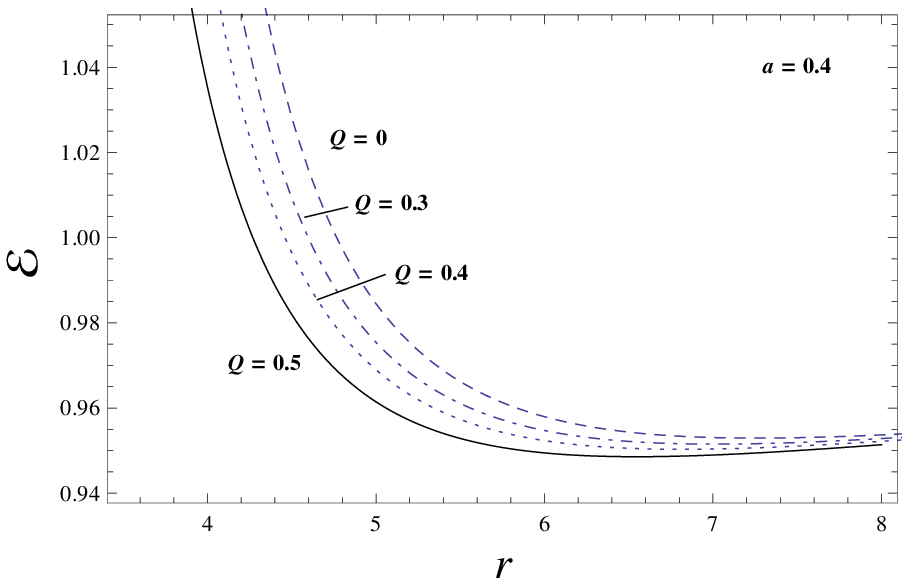}
\includegraphics[width=0.45\linewidth]{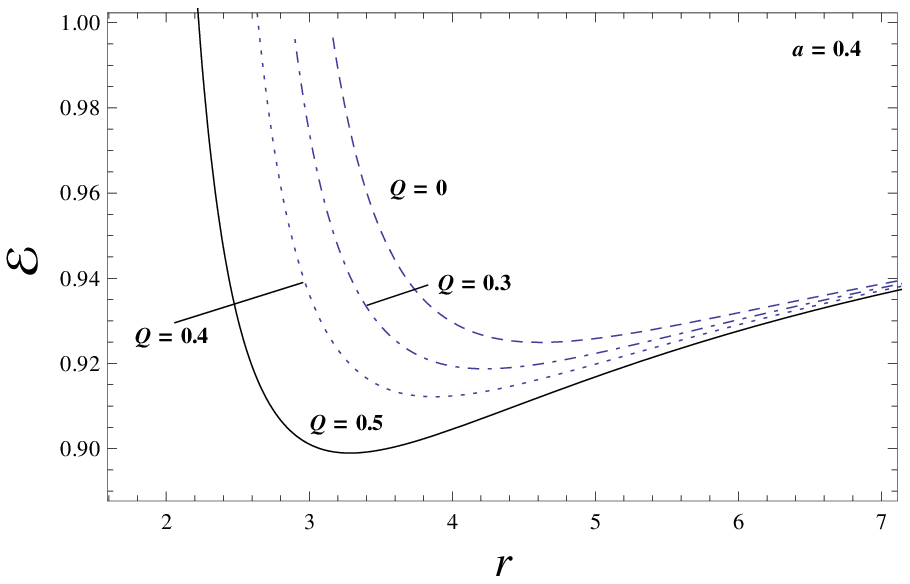}

\includegraphics[width=0.45\linewidth]{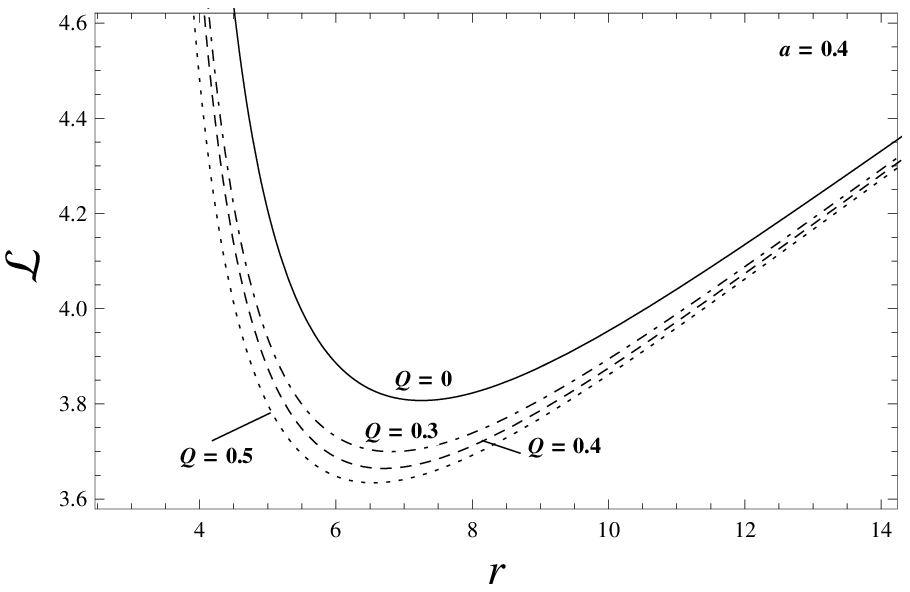}
\includegraphics[width=0.45\linewidth]{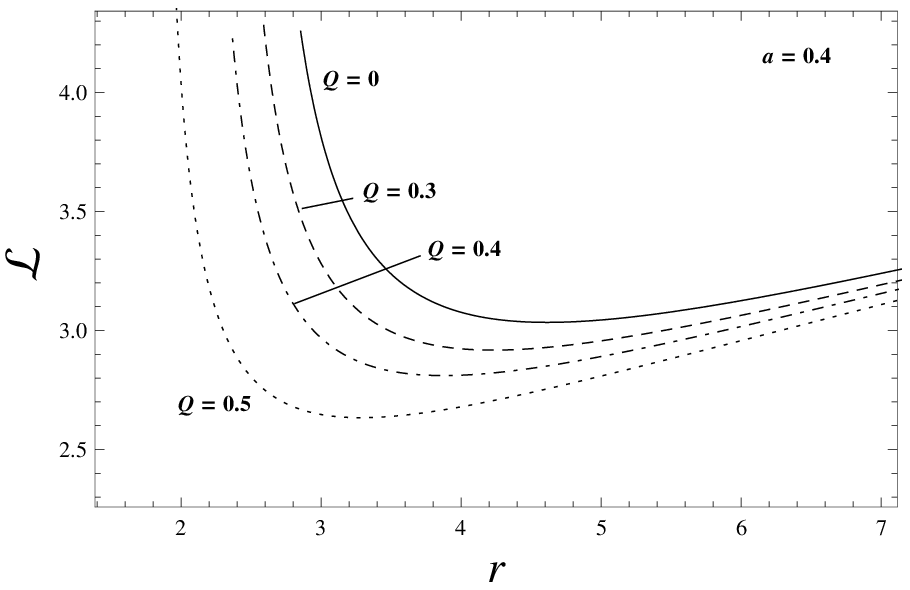}
\end{center}
\caption{\label{fig_bobir4} Radial dependence of the specific
energy and angular momentum  of co-rotating (left, top and bottom)
and counter rotating (right, top and bottom) particle at the
circular geodesics around   black hole for the different values of
the electric charge $Q$ from left to right, respectively.}
\end{figure*}

\begin{figure*}[t!.]
\begin{center}
\includegraphics[width=0.45\linewidth]{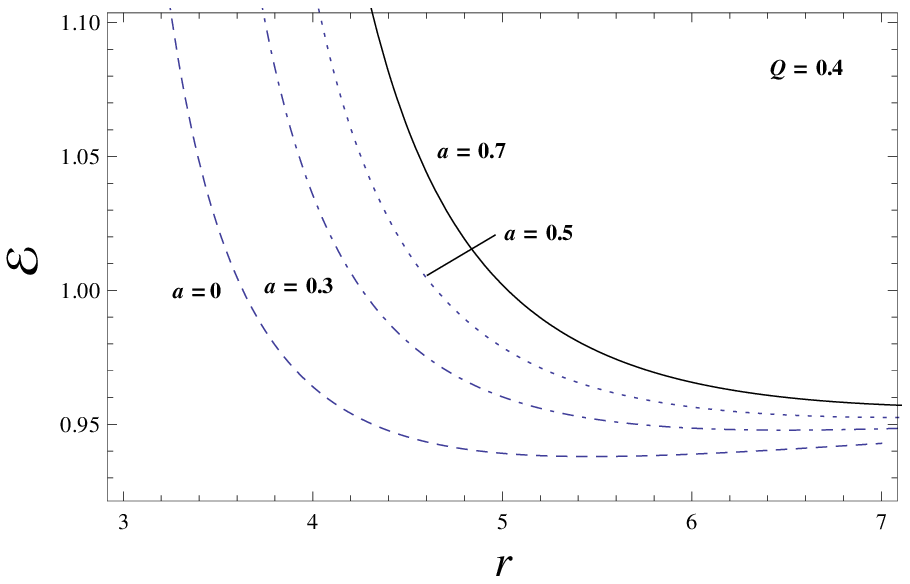}
\includegraphics[width=0.45\linewidth]{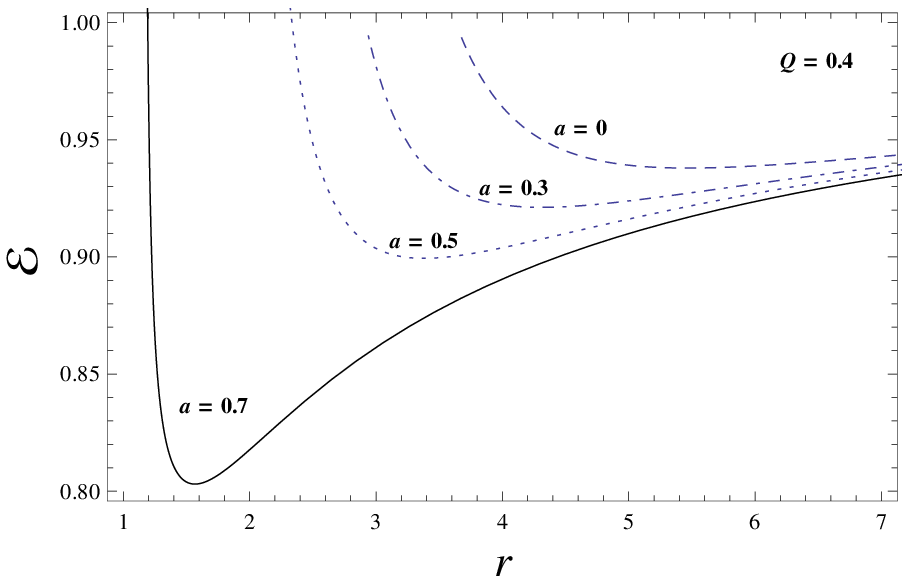}

\includegraphics[width=0.45\linewidth]{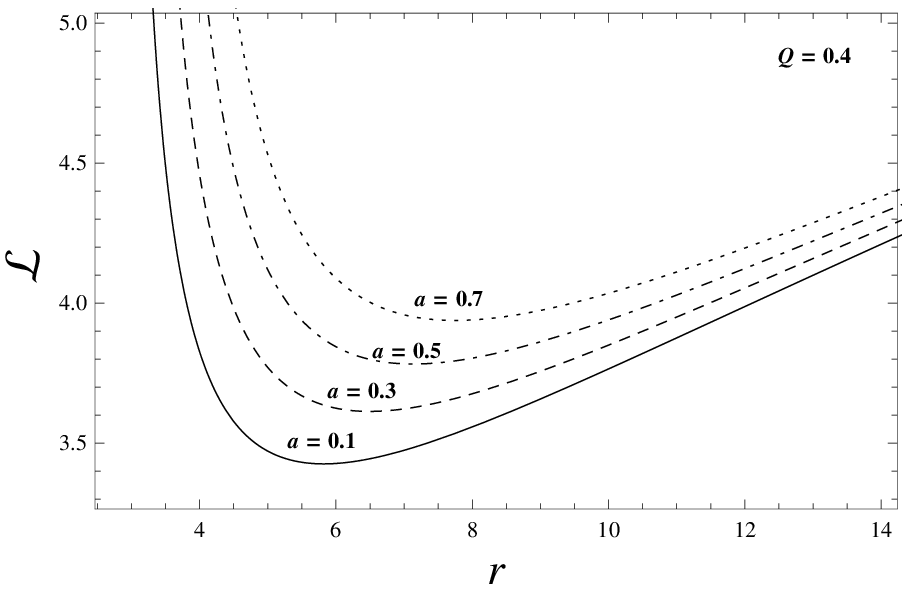}
\includegraphics[width=0.45\linewidth]{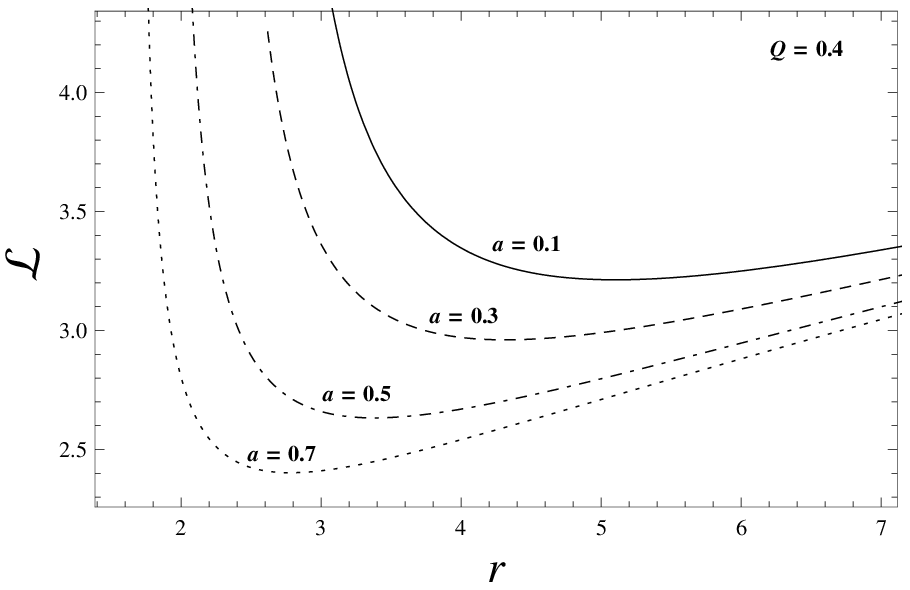}
\end{center}
\caption{\label{fig_bobir5} Radial dependence of the specific
energy and angular momentum  of co-rotating (left, top and bottom)
and counter rotating (right, top and bottom) particle at the
circular geodesics around   black hole for the different values of
the rotation parameter $a$ from left to right, respectively.}
\end{figure*}

\begin{figure*}[t!.]
\begin{center}
\includegraphics[width=0.24\linewidth]{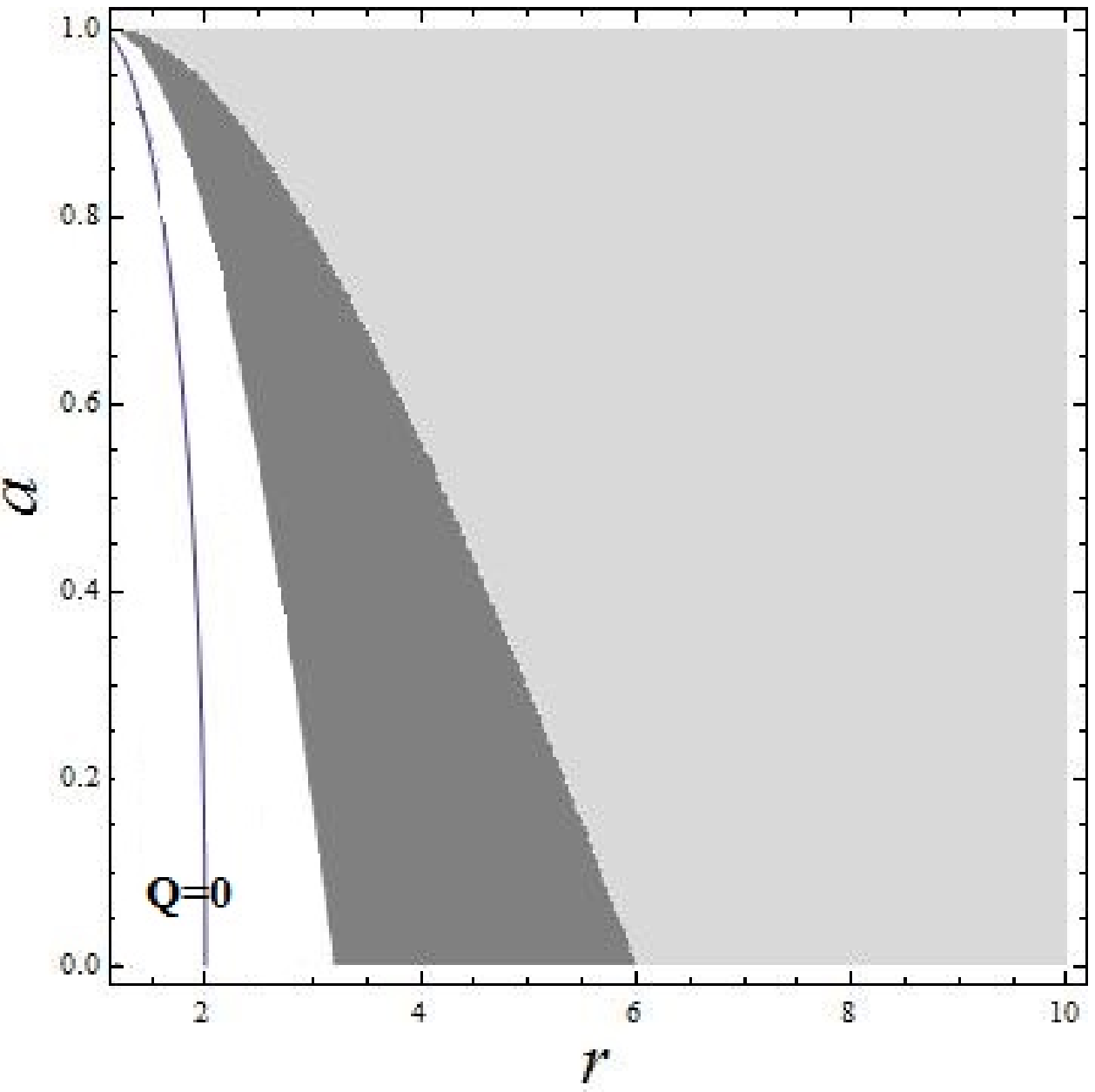}
\includegraphics[width=0.24\linewidth]{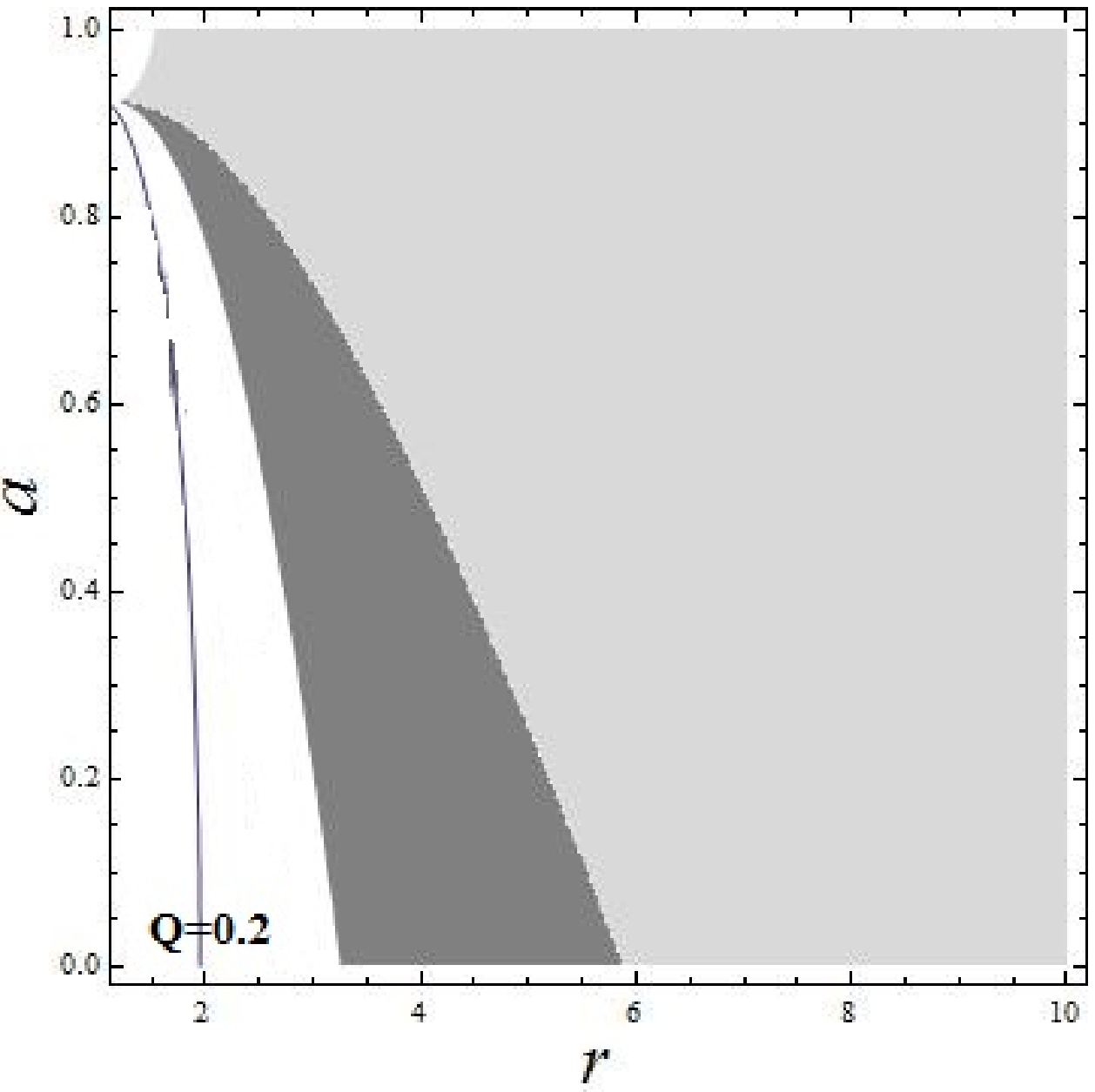}
\includegraphics[width=0.24\linewidth]{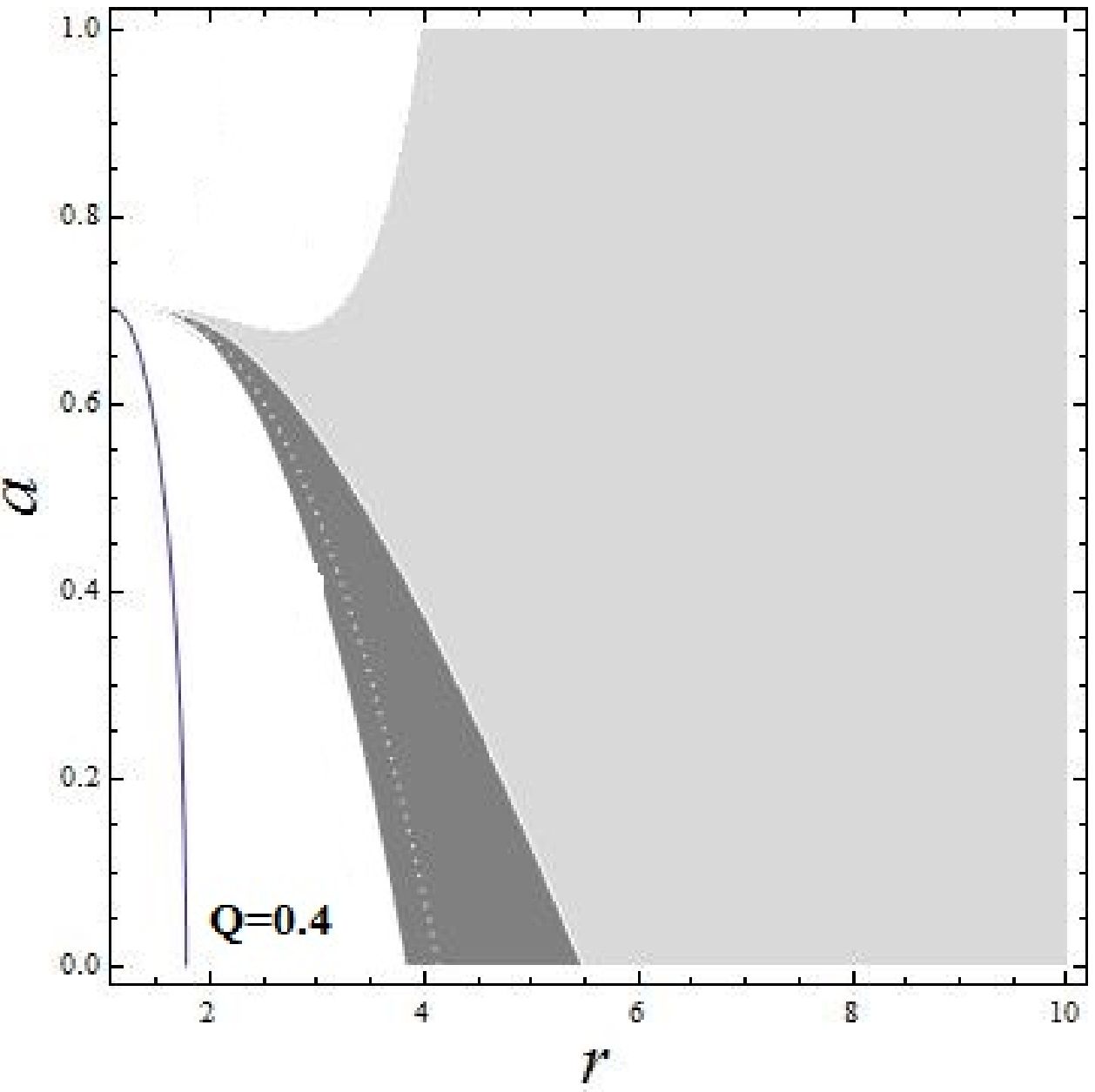}
\includegraphics[width=0.24\linewidth]{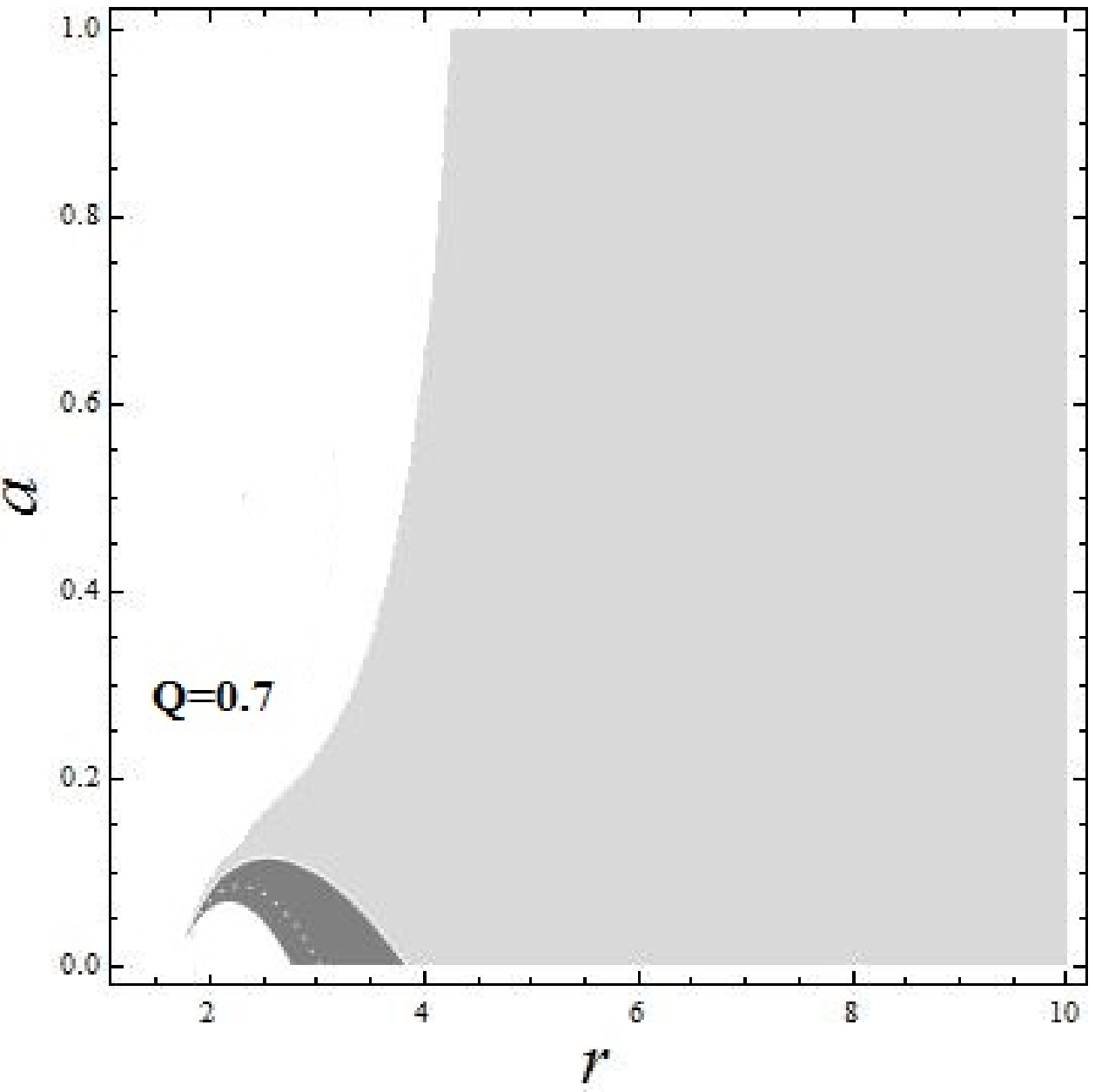}
\end{center}
\caption{\label{fig_bobir3} The regions of the circular geodesics
of the particle around rotating regular   BH for the different
values of the electric charge $Q$. Gray and light gray regions
represent unstable and stable orbits, respectively. Solid line
represents event horizon of the   BH.}
\end{figure*}

As already pointed out in the previous section,  when $Q=0$ the
spacetime metric (\ref{00}) coincides with the Kerr one and $Q=0$,
$a=0$ with Schwarzschild one. From the first graph of
Fig.~\ref{fig_bobir3} one can see the following remarks:

\begin{itemize}

\item {In the case of extreme BH (inner and outer horizons merge into one) the radius of ISCO coincides with one of the event horizon of the BH.}

\item When $Q=0$, $a=0$ the spacetime metric (\ref{00}) {is identical to} the Schwarzschild one and the radius of ISCO is $r=6$ ($M=1$).

\item When $Q=0$ the spacetime metric (\ref{00})
{red}{is identical to} the Kerr one. The radius of ISCO of the extreme ($a=1$) Kerr BH is $r=1$ ($M=1$).

\end{itemize}

\section{Center of mass energy of particles in collision}
\label{bsw}

Now based on the BSW (Ba\v{n}ados-Silk-West) mechanism  of the
energy extraction from the rotating black hole we calculate center
of mass energy $E_{CM}$ for collision of the two neutral identical
particles with mass $m_{1}=m_{2}=m_{0}$. We assume that particles
are coming from infinity with $E_{1}/m_{0}=E_{2}/m_{0}=1$ and
approaching the black hole with the different angular momenta
$L_{1}$ and $L_{2}$ as well as the particles motion and their
collisions occur in the equatorial plane $\theta=\pi/2$.

The center of mass energy can be found by using the following
formula~\cite{Banados}:
\begin{eqnarray}\label{40}
\frac{E_{CM}^2}{2m_{0}^2}=1-g_{\mu\nu}u_{1}^{\mu}u_{2}^{\nu}\ ,
\end{eqnarray}
where $u_{1}^{\mu}$ and $u_{2}^{\nu}$ are four velocities  of the
first and second particles, respectively. The four velocity of the
particle that is moving around rotating   black hole in the
equatorial plane is given by the expressions~(\ref{08})
and~(\ref{005}). For simplicity, considering ${\cal E}_1={\cal
E}_2=1$ and inserting the expressions~(\ref{08}) and~(\ref{005})
into~(\ref{40}), we get the center of mass energy as
\begin{eqnarray}\label{41}
&&\frac{E_{CM}^2}{2m_{0}^2}=\frac{1}{(r^2+a^2)(r^2+Q^2)^2-r^4(-Q^2+2\sqrt{r^2+Q^2})}\nonumber\\
&&\left\{-a r^2({\cal L}_1+{\cal L}_2)(-Q^2+2\sqrt{r^2+Q^2})-{\cal L}_1{\cal L}_2[(r^2+Q^2)^2\right.\nonumber\\
&&\left.-r^2(-Q^2+2\sqrt{r^2+Q^2})]+2(r^2+a^2)(r^2+Q^2)^2\right.\nonumber\\
&&\left.-r^2(r^2-a^2)(-Q^2+2\sqrt{r^2+Q^2})-\sqrt{R_{1}}\sqrt{R_{2}}\right\},
\end{eqnarray}
where
\begin{eqnarray}\label{42}
\begin{split}
R_{i}(r)&=r^2(-Q^2+2\sqrt{r^2+Q^2})(a-{\cal L}_i)^2-{\cal L}_i^2(r^2+Q^2)^2\\
&+r^4(-Q^2+2\sqrt{r^2+Q^2}), \hspace{1.5cm} i=1,2.
\end{split}
\end{eqnarray}
{In absence of the electric charge $Q=0$ the
expressions~(\ref{41}) and~(\ref{42}) will reduce to the ones for
the Kerr black hole~\cite{Banados}.}
\begin{figure*}[t!.]
\begin{center}
\includegraphics[width=0.49\linewidth]{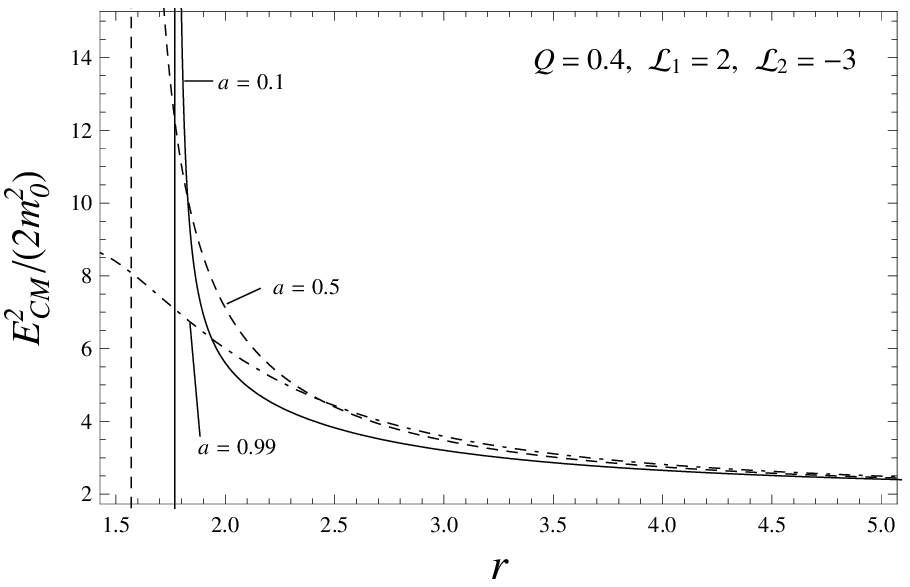}
\includegraphics[width=0.49\linewidth]{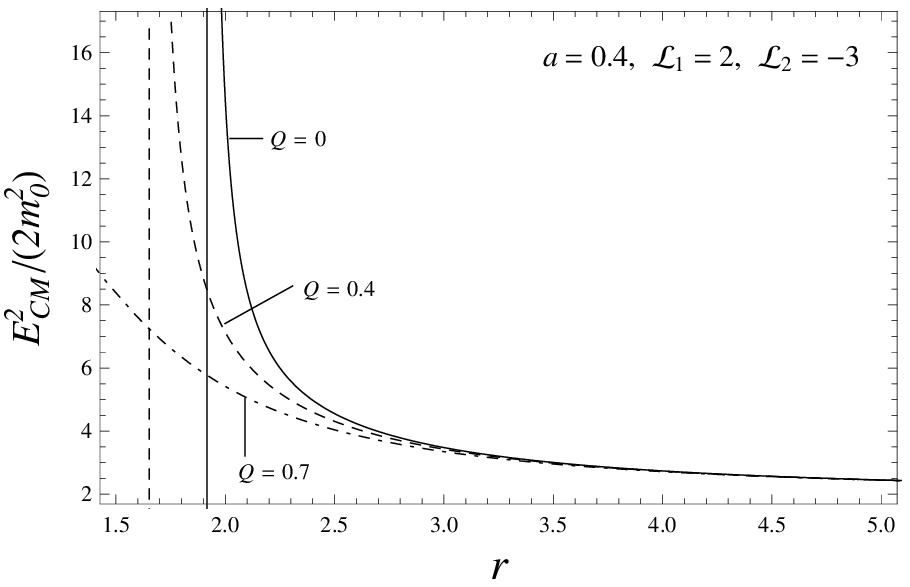}

\includegraphics[width=0.49\linewidth]{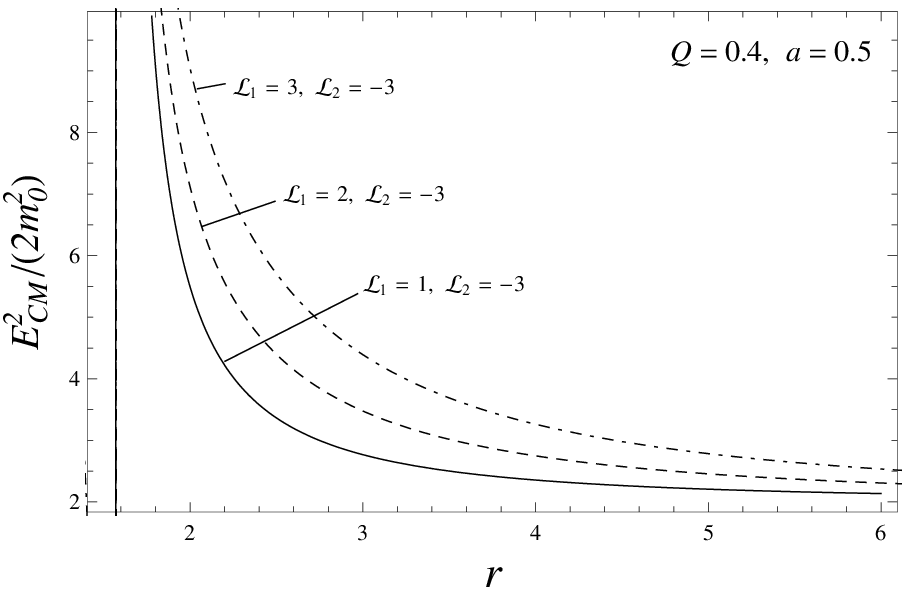}
\includegraphics[width=0.49\linewidth]{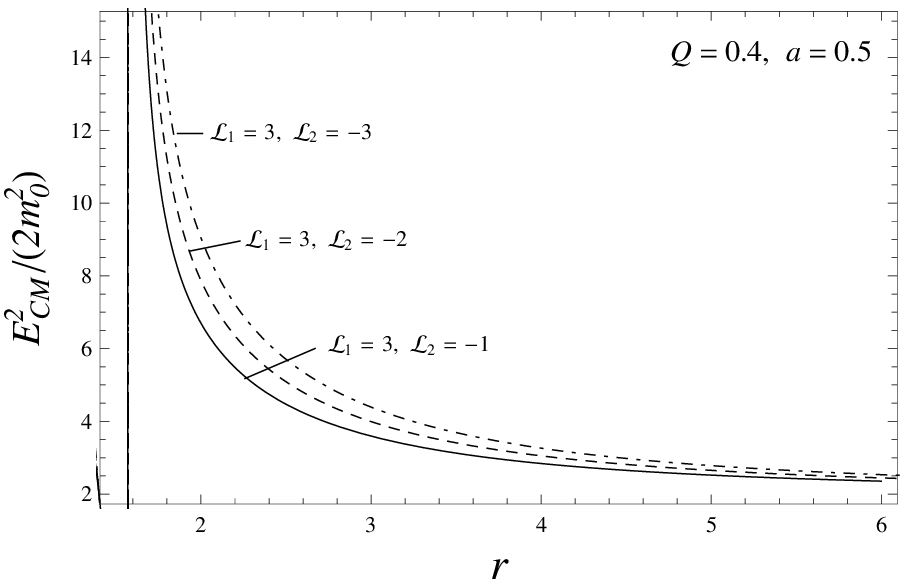}
\end{center}
\caption{\label{fig_ahmadjon3} Radial dependence  of the center of
mass energy for few typical values of the rotation parameter $a$,
electric charge $Q$ and angular momenta of the particles ${\cal
L}_1$ and ${\cal L}_2$. {Vertical line represents the location of
event horizon of the BH.}}
\end{figure*}

The horizon of the   black hole is located at
$(r^2+a^2)(r^2+Q^2)^2-r^4(-Q^2+2\sqrt{r^2+Q^2})=0$.  Therefore, at
the one sight it seems that the center of mass energy diverges at
the horizon of the   black hole. However, at this point numerator
of the~(\ref{41}) diverges also and in order to eliminate this
uncertainty one can use L'Hopital's rule.

{At the next step  we compute  the maximal value of energy which
can be extracted through the acceleration process~ from the
rotating regular black hole. For this we first calculate the
energy of a test particle moving along the innermost stable
circular orbit. Then we use the definition of the coefficient of
total amount of released energy of the test particle shifting from
the outward stable circular orbit with the radius  $r_c$ to the
innermost stable circular orbit. Then coefficient of the energy
release efficiency can be found as
\begin{eqnarray}
\eta_c=100\times \frac{E(r_{c})-E(r_{ISCO})}{E(r_{c})} \ .
\end{eqnarray}}
%
%%%%%%%%%%%%%%%%%
%

{The efficiency coefficient $\eta$ for the different values of the
rotation parameter $a$ and electric charge $Q$ is shown in the
Table~\ref{efficiency}. The energy extraction is essentially
amplified with the increase of the electric charge of the rotating
regular black hole. In the limiting case when the extreme rotating
black hole is uncharged one has the maximal efficiency $42\%$. For
the rotating black hole with smaller rotation parameter $a$ the
presence of the electric charge $Q$ fulfils the effect of rotation
of the black hole and increases the energy extraction efficiency.
Physically this means that the electric charge decreases the
potential of the gravitational field and particle needs less bound
energy at the circular geodesics. In the case when the electric
charge destroys the singularity at the origin and no horizon for
the black hole the circular geodesics can exist at any radial
distance from the central object with sufficiently small bound
energy. }
\begin{table}[hbt]
\begin{center}
\caption{\label{efficiency} The efficiency of the energy
extraction  $\eta_c$ (\%) from the   black hole by the particle
acceleration mechanism for several values of the rotation
parameter $a$ and charge $Q$.}

\begin{tabular}{c c c c c c c c}
\hline
\hline
$a$& $Q=0 $ & $0.1$ & $0.2$ & $0.3$ & $0.4$ & $0.5$  & $0.6$  \\[0.8ex]
\hline
$0 $ & 5.75\  & 6.04\  & 6.40\  & 6.85\  & 7.48\  & 8.51 & 9.78\ \\[0.8ex]
\hline
$0.1 $ & 6.06\  & 6.41\  & 6.84\  & 7.42\  & 8.28\  & 9.72\ &  11.62\ \\[0.8ex]
\hline
$0.2 $ & 6.46\  & 6.89\  & 7.44\  & 8.23\  & 9.66\  &  11.61\ &  13.64\  \\[0.8ex]
\hline
$0.3 $ & 6.94\  & 7.47\  & 8.21\  & 9.41\  & 11.56\  &  13.82\ &  \\[0.8ex]
\hline
$0.4 $ & 7.51\  & 8.21\  & 9.27\  & 11.5\  &  13.76\ & 14.12\  & \\[0.8ex]
\hline
$0.5 $ & 8.21\  & 9.18\  & 10.9\  & 12.54\ & 13.98\  &  15.01\ &  \\[0.8ex]
\hline
$0.6 $ & 9.12\  & 10.58\  & 14.44\  & 17.1\  &  19.32\ &  &  \\[0.8ex]
\hline
$0.7 $ & 10.36\  & 12.92\  & 18.14\   & 21.17\ &  18.12\ &  &  \\[0.8ex]
\hline
$0.8 $ & 12.21\  & 20.14\  &  22.2\ & 29.12\  &  &  &  \\[0.8ex]
\hline
$0.9 $ & 15.58\  & 21.12\  & 27.31\ &  &  &   \\[0.8ex]
\hline
$1 $ & 42\  &  &  &  &  &   \\[0.8ex]
\hline
\hline
\end{tabular}
\end{center}
\end{table}

\section{Energy extraction from rotating regular black hole by Penrose process}
\label{penrose}

We have already known from \cite{Ghosh} that the Penrose process
is  one theorized by Roger Penrose in 1969 wherein energy can be
extracted from a rotating black hole. Energy extraction occurs not
inside the event horizon of the black hole, it occurs in the
region of ergosphere on account of rotational energy of the black
hole. In this process massive particle enters into the ergosphere
and splits into two pieces: one of them escapes from the black
hole to infinity while the other one falls into the black hole.
The escaping piece can possibly have greater energy than the
infalling one, due to the infalling piece has negative energy. As
a result of this process black hole reduces its angular momentum
and consequently energy of the black hole is extracted. It derives
from signature of energies of two pieces that the escaping
particle has more energy than one which entered into ergosphere.

Assume a particle enters into ergosphere of the  black hole and is
splitted into two labeled as $1$ and $2$ pieces. The first piece
$1$ has more energy $(E_{1})$ than the incident particle 0 and
exits ergosphere while the second piece $2$ is falling into the
black hole with negative energy $E_{2}$ \cite{Nozawa}, i.e.
according to the law of conservation of energy
\begin{eqnarray}\label{15}
E_{0}=E_{1}+E_{2}\ ,
\end{eqnarray}
where $E_{2}<0$, then $E_{1}>E_{0}$. For simplicity we assume that
all particles are confined on the equatorial plane
($\theta=\pi/2$)
\begin{eqnarray}\label{16}
\upsilon=\frac{dr}{dt}\ ,\ \ \  \Omega=\frac{d\phi}{dt}\ ,
\end{eqnarray}
where $\upsilon$ and $\Omega$ are the radial and angular velocity
of the particle with respect to an observer at asymptotic
infinity.

It is known that in the Penrose process energy of rotating black
hole is extracted on account of decreasing black holes angular
momentum. From the conservation laws of energy and angular
momentum we have
\begin{eqnarray}\label{17}
E=-p^t A\ , \ L=p^t \Omega\ ,   \ A\equiv g_{tt}+\Omega g_{t\phi}\
.
\end{eqnarray}
From the Hamilton-Jacobi equation for the timelike geodesics,
namely $p^\mu p_\mu=-m^2$, one can obtain
\begin{eqnarray}\label{18}
g_{tt}\dot{t}^2+g_{rr}\dot{r}^2+g_{\phi\phi}\dot{\phi}^2+2g_{t\phi}\dot{t}\dot{\phi}=-m^2\
.
\end{eqnarray}
Dividing both sides of (\ref{18}) by $\dot{t}^2$ and using
(\ref{16}) and (\ref{17}) one can get
\begin{eqnarray}\label{19}
g_{tt}+g_{rr}
\upsilon^2+g_{\phi\phi}\Omega^2+2g_{t\phi}\Omega=-m^2(\frac{A}{E})^2\
.
\end{eqnarray}
As one can see the right hand side of the expression (\ref{19}) is
negative or equals to zero and the second term in the left hand
side of the expression (\ref{19}) is always positive. Due to this
one can write the expression (\ref{19}) in the following form
\cite{Nozawa}:
\begin{eqnarray}\label{20}
g_{\phi\phi}\Omega^2+2g_{t\phi}\Omega+g_{tt}=-m^2(\frac{A}{E})^2-g_{rr}
\upsilon^2\leq0\ .
\end{eqnarray}

From the inequality (\ref{20}) it follows that the value of
$\Omega$ is in the range of $\Omega^-\leq\Omega\leq\Omega^+$
\cite{Nozawa}. Here $\Omega^\pm$ is
\begin{equation}\label{21}
\Omega^\pm=-\frac{g_{t\phi}}{g_{\phi\phi}}\pm\sqrt{\frac{g_{t\phi}^2}{g_{\phi\phi}^2}-\frac{g_{tt}}{g_{\phi\phi}}}\
.
\end{equation}
Using the expression (\ref{17}) the equations of the conservation
of energy (\ref{15}) and angular momentum can be written as
\cite{Nozawa}
\begin{eqnarray}\label{22}
p_{(0)}^t A_{(0)}=p_{(1)}^t A_{(1)}+p_{(2)}^t A_{(2)}\ ,
\end{eqnarray}
\begin{eqnarray}\label{23}
p_{(0)}^t\Omega_{(0)}=p_{(1)}^t \Omega_{(1)}+p_{(2)}^t
\Omega_{(2)}\ .
\end{eqnarray}

The energy extraction from black holes and its efficiency is the
momentous problem of the general relativity, see e.g. \cite{Liu}
for its discussion. There are several means and processes which
 are dedicated to the determination of the efficiency of the energy extraction from the rotating black holes.
 One of these processes is the Penrose one and  one can obtain its
 efficiency
\begin{eqnarray}\label{24}
\eta=\frac{|E_{(2)}|}{E_{(0)}}=\frac{E_{(1)}-E_{(0)}}{E_{(0)}}=\chi-1\
,
\end{eqnarray}
using the expression (\ref{15}) and taking into account
$E_{(2)}<0$ \cite{Nozawa} where $\chi=E_{(1)}/E_{(0)}$\ and
$\chi>1$. With the help of the expressions (\ref{17}), (\ref{22})
and (\ref{23})
\begin{eqnarray}\label{25}
\chi=\frac{E_{(1)}}{E_{(0)}}=\frac{(\Omega_{(0)}-\Omega_{(2)})A_{(1)}}{(\Omega_{(1)}-\Omega_{(2)})A_{(0)}}\
.
\end{eqnarray}

Following to~\cite{Nozawa} assume that the incident particle has
initial energy $E_{(0)}=1$ and is splitted into a pair of two
photons in the black hole ergosphere, namely their momenta are
equal to zero ($p_{(1)}=p_{(2)}=0$). As one can  see from the
expression (\ref{25}) that the maximum value of the efficiency of
the Penrose process in this case corresponds to the maximum value
of $\Omega_{(2)}$ and the minimum value of $\Omega_{(1)}$ at the
same time. At this moment the radial velocities of both pieces
will vanish ($\upsilon_{(1)}=\upsilon_{(2)}=0$), namely
\begin{eqnarray}\label{26}
\Omega_{(1)}=\Omega^+,\nonumber\\
\Omega_{(2)}=\Omega^-,
\end{eqnarray}
and the corresponding values of the parameter $A$ are
\begin{eqnarray}\label{26.1}
A_{(0)}=g_{tt}+\Omega_{(0)}g_{t\phi}\ , \
A_{(2)}=g_{tt}+\Omega^-g_{t\phi}\ .
\end{eqnarray}
Consequently, the four momenta of the pieces are \cite{Liu}
\begin{eqnarray}\label{27}
p_{i}=p^{t}(1,0,0,\Omega_{i}), \ i=1,2\ .
\end{eqnarray}
Due to the zero radial velocity $\upsilon_{(i)}=0$, the equation
(\ref{19}) takes a form
\begin{eqnarray}\label{28}
(g_{\phi\phi}+g_{t\phi}^2)\Omega^2+2g_{t\phi}(1+g_{tt})\Omega+g_{tt}(1+g_{tt})=0\
.
\end{eqnarray}
Angular velocity of the incident particle can be derived from
(\ref{28}) as~\cite{Nozawa}
\begin{eqnarray}\label{29}
\Omega_{(0)}=\frac{-g_{t\phi}(1+g_{tt})+\sqrt{(1+g_{tt})(g_{t\phi}^2-g_{tt}g_{\phi\phi})}}{g_{\phi\phi}+g_{t\phi}^2}\
.
\end{eqnarray}
Putting the expressions (\ref{26}) and (\ref{26.1}) into
(\ref{25})  we have the expression for the efficiency of the
energy extraction as~\cite{Nozawa}
\begin{eqnarray}\label{30}
\eta=\frac{(\Omega_{(0)}-\Omega^-)(g_{tt}+\Omega^+g_{t\phi})}{(\Omega^+-\Omega^-)(g_{tt}+\Omega_{(0)}g_{t\phi})}-1\
.
\end{eqnarray}
In order to achieve the maximum  value of the efficiency the
incident particle must be splitted into two pieces at the horizon
of the black hole \cite{Liu} and in this case the expression
(\ref{30}) takes a form
\begin{eqnarray}\label{31}
\eta_{max}=\frac{\sqrt{1+g_{tt}}-1}{2}|_{r=r_{+}}\ .
\end{eqnarray}

In the Table~\ref{tab1} the values of the maximum efficiency of
the energy extraction from the regular black hole by the Penrose
process is given for the several typical values of the rotation
parameter $a$ and the electric charge $Q$. According to the
Table~\ref{tab1} the maximum value of the efficiency of the energy
extraction from the   black hole is smaller than one from the Kerr
black hole and when the ergosphere of the black hole vanishes the
energy extraction does not occur.

\begin{table}[hbt]
\begin{center}
\caption{\label{tab1} The values  of the maximum efficiency of the
energy extraction $\eta_{max}$ (\%) from the   black hole by the
Penrose process for several values of the rotation parameter $a$
and charge $Q$.}

\begin{tabular}{c c c c c c c c}
\hline
\hline
$a$& $Q=0 $ & $0.1$ & $0.2$ & $0.3$ & $0.4$ & $0.5$ & $0.6$ \\[0.8ex]
\hline
$0.1 $ & 0.06\  & 0.06\  & 0.06\  & 0.07\  & 0.08\  & 0.10\  & 0.15\  \\[0.8ex]
\hline
$0.2 $ & 0.25\  & 0.26\  & 0.27\  & 0.29\  & 0.33\  & 0.40\  & 0.66\  \\[0.8ex]
\hline
$0.3 $ & 0.59\  & 0.60\  & 0.62\  & 0.67\  & 0.76\  & 0.96\   &  \\[0.8ex]
\hline
$0.4 $ & 1.08\  & 1.09\  & 1.14\  & 1.24\  & 1.44\  & 1.91\   & \\[0.8ex]
\hline
$0.5 $ & 1.77\  & 1.80\  & 1.90\  & 2.07\  & 2.47\  & 3.80\  &  \\[0.8ex]
\hline
$0.6 $ & 2.70\  & 2.75\  & 2.92\  & 3.28\  & 4.13\  &  &  \\[0.8ex]
\hline
$0.7 $ & 4.01\  & 4.10\  & 4.41\  & 5.13\  & 8.82\  &  &  \\[0.8ex]
\hline
$0.8 $ & 5.90\  & 6.08\  & 6.73\  & 8.86\  &  &  &  \\[0.8ex]
\hline
$0.9 $ & 9.01\  & 9.45\  & 11.75\  &  &  &  &  \\[0.8ex]
\hline
$1.0 $ & 20.71\  &  &  &  &  &  &  \\[0.8ex]
\hline
\hline
\end{tabular}
\end{center}
\end{table}

One of consequences of the energy extraction  from the black hole
is irreducible mass of the black hole. As a result of a big number
of infalling particles into the black hole with negative energy
the mass of the black hole changes by $\delta M=E$
~\cite{Abdujabbarov}. There is no upper limit on change of the
mass of the black hole. However, each infalling particle with
negative energy decreases the mass of the black hole untill its
irreducible mass. This is why there is a lower limit on the mass
of the black hole.

In order to find the lower limit $\delta M$ we rewrite (\ref{07})
with help of the expressions (\ref{04}) and (\ref{05}) in the
following form
\begin{eqnarray}\label{35}
\alpha E^2+2\beta E+\gamma+g_{rr}p_{r}^2+m^2=0\ ,
\end{eqnarray}
where
\begin{eqnarray}\label{36}
\alpha=-\frac{g_{\phi\phi}}{g_{t\phi}^2-g_{tt}g_{\phi\phi}},\\
\beta=-\frac{g_{t\phi}L}{g_{t\phi}^2-g_{tt}g_{\phi\phi}},\\
\gamma=-\frac{g_{tt}L^2}{g_{t\phi}^2-g_{tt}g_{\phi\phi}}\ .
\end{eqnarray}

Assuming  that at the horizon $p_{r}=0$, $m=0$ and solving the
equation (\ref{35}) with respect to $E$ we get
\begin{eqnarray}\label{37}
E=-\frac{\beta}{\alpha} \pm
\sqrt{\frac{\beta^2-\alpha\gamma}{\alpha^2}}\ .
\end{eqnarray}

At the horizon $r=r_{+}$ the discriminant of the equation
(\ref{37}) is equal to zero and consequently the lower limit of
$\delta M$ is
\begin{eqnarray}\label{38}
\delta M=\frac{a L}{r_{+}^2+a^2}.
\end{eqnarray}

The derived limit (\ref{38}) formally coincides with the
expression derived for the Kerr black but in reality it is
different and more massive due to the different value of $r_{+}$
for the regular   black hole.

\section{Conclusion}
\label{conclusion}

In this paper we have studied the neutral particle  motion and the
energy extraction from the regular   black hole. The dependence of
the ISCO (innermost stable circular orbits along geodesics) and
unstable orbits on the value of the electric charge of the
rotating regular black hole is studied. In particular we have
shown that with the increase of the value of the electric charge
$Q$ the radius of the ISCO decreases.

Energy extraction from the rotating regular black hole through the
different processes has been examined. We have found expression of
the center of mass energy for the colliding neutral particles
coming from infinity, based on the BSW (Ba\v{n}ados-Silk-West)
mechanism. In particular we have calculated the center-of-mass
frame energy by colliding two neutral particles of the same mass
parameter around rotating regular   black hole.  It has been shown
that two colliding neutral particles which are at rest infinity
with different angular momentums can give arbitrarily large value
of the center of mass energy. {The electric charge $Q$ of rotating
regular black hole decreases the potential of the gravitational
field and the particle needs less bound energy at the circular
geodesics. This causes the increase of efficiency of the energy
extraction  through BSW process in the presence of the electric
charge $Q$ from rotating regular black hole.}

Efficiency of the energy extraction by the Penrose  process from
the   black hole has been calculated. It has been shown that the
efficiency of the energy extraction from the rotating regular
black hole via the Penrose process decreases with the increase of
the electric charge $Q$ and is smaller with compare to 20.7~\%
which is the value for the extreme Kerr black hole with the
specific angular momentum $a=1$. It is due to the fact that on
account of the nonvanishing electric charge $Q$ the ergosphere of
the   black hole decreases and for the limiting value of the
electric charge $Q>0.634$ ergoregion vanishes. After vanishing of
the ergosphere the energy extraction does not occur.

\begin{acknowledgments}

The~authors acknowledge the~project Supporting Integration  with
the~International Theoretical and Observational Research Network
in Relativistic Astrophysics of Compact Objects, Grant No.
CZ.1.07/2.3.00/20.0071, supported by Operational Programme
Education for Competitiveness funded by the Structural Funds of
the~European Union. Z.S. acknowledges the Albert Einstein Center
for gravitation and astrophysics supported by the~Czech Science
Foundation Grant No.~14-37086G. A.A. and B.A. thank the Goethe
University, Frankfurt am Main, Germany, and the Faculty of
Philosophy and Science, Silesian University in Opava, Czech
Republic, for the warm hospitality. This research is supported in
part by Grants No. F2-FA-F113, No. EF2-FA-0-12477, and No.
F2-FA-F029 of the UzAS, and by the ICTP through Grant No.
OEA-PRJ-29 and No. OEA-NET-76 and by the Volkswagen Stiftung,
Grant No. 866.

\end{acknowledgments}

\label{lastpage}

\end{document}